\RequirePackage{fix-cm}

\documentclass[twocolumn]{svjour3}        
\smartqed  
\usepackage{graphicx}

\newcommand{\beq}{\begin{equation}}
\newcommand{\eeq}{\end{equation}}
\newcommand{\beqa}{\begin{eqnarray}}
\newcommand{\eeqa}{\end{eqnarray}}
\newcommand{\ba}{\begin{array}}
\newcommand{\ea}{\end{array}}

\begin{document}

\title{Quantum fluctuations and vortex-antivortex unbinding \\ 
in the 2D BCS-BEC crossover}

\author{L. Salasnich and G. Bighin}

\institute{L. Salasnich \at
              Dipartimento di Fisica e Astronomia ``Galileo Galilei'', \\
Universita di Padova, Italy  \\
INO-CNR, Sezione di Sesto Fiorentino, Italy \\
              \email{luca.salasnich@unipd.it} \\
G. Bighin \at
Dipartimento di Fisica e Astronomia ``Galileo Galilei'', \\
Universita di Padova, Italy  \\
\email{bighin@pd.infn.it} }

\date{Received: date / Accepted: date}

\maketitle

\begin{abstract}
Very recently quasi two-dimensional (2D) systems made of attractive fermionic 
alkali-metal atoms with a widely tunable interaction 
due to Fano-Feshbach resonances have been realized. 
In this way it has been achieved the 2D crossover 
from the Bardeen-Cooper-Schrieffer regime of weakly-interacting 
Cooper pairs to the Bose-Einstein condensate regime strongly 
bound dimers. These experiments pave the way to the investigation of 
2D strongly-interacting attractive fermions during the 
Berezinskii-Kosterlitz-Thouless (BKT) transition from a low-temperature 
superfluid phase characterized by quasi-condensation to a 
high-temperature normal phase, where vortex proliferation driven 
by quantum and thermal fluctuations 
completely destroys superfluidity. In this paper we discuss 
our preliminar theoretical results 
on the behavior of the BKT critical temperature across the crossover. 
Our microscopic calculations are based on functional integration 
taking into account renormalized Gaussian fluctuations and 
the crucial 2D effect of vortex-antivortex unbinding.

\keywords{BCS-BEC crossover \and Ultracold atoms 
\and Dimensional regularization}
\end{abstract}

\section{Introduction} 
 
The study of superfluid properties of Fermi systems is one of the most 
interesting areas of research in the field of ultracold atoms because 
of direct implications for superconductivity in solid-state materials 
as well as for nuclear matter and other many-body quantum systems 
\cite{[1],[2]}. 
An extremely important achievement in this field has been the realization 
of the crossover from the Bardeen-Cooper-Schrieffer (BCS) superfluid 
phase of loosely bound pairs of fermions to the Bose-Einstein condensate 
(BEC) of tightly bound composite bosons \cite{[3]}. The actual experimental 
realization of the crossover was made possible by the use of 
Fano-Feshbach resonances, a tool from atomic physics, with no counterpart 
in solid-state experiments, which allows to change at will the 
strength and sign of the interparticle interactions \cite{[3]}. 

Several exciting research lines are now focused on Fermi superfluids 
in low spatial dimensions, where quantum fluctuations are strongly enhanced 
and beyond-mean-field theories are needed to obtain a reliable description 
of the observed phenomena. While for one-dimensional (1D) bosonic and 
fermionic systems with contact interaction there are exact analytical 
solutions \cite{[32],[33],[34],[35],[36],[37],[38],[39]}, for 
two-dimensional (2D) quantum systems there are 
no exact solutions. In 2D systems with a continuous symmetry the 
Mermin-Wagner-Hohenberg theorem precludes spontaneous symmetry breaking 
and condensation at finite temperature \cite{[1],[2]}. 
The Berezinskii-Kosterlitz-Thouless critical temperature marks 
the transition from a superfluid phase 
characterized by quasi-condensation and algebraic long-range order to 
a normal phase, where vortex proliferation completely destroys superfluidity. 
As opposed to conventional off-diagonal long-range order typical of 
3D superfluid systems, algebraic long-range order is driven by strongly 
enhanced quantum and thermal fluctuations \cite{[1],[2]}. 
For this reason of particular relevance are studies aimed at understanding 
the pairing of fermions in strongly interacting 2D Fermi gases, both 
in the homogeneous case and in the presence of an optical lattice.  
This subject is of great importance for condensed matter physics also in 
view of the not yet fully understood character of the corresponding 
mechanisms in high-temperature superconductors \cite{[40]}. Strongly 
interacting Fermi gases made of alkali-metal atoms 
in quasi-2D configurations have been realized only very recently 
\cite{[41],[42],[43],[44]} and some evidences of superfluid behavior, 
such as the condensation of pairs \cite{[42]} and the algebraic 
decay of the first-order correlation function \cite{[43]} have been 
observed. However, genuine signatures of the superfluid state 
such as the second-sound mode are still missing. 

\section{Results at zero temperature} 

The first theoretical analysis of 2D Fermi superfluid in the full 
BCS-BEC crossover based on Gaussian corrections to the mean-field 
equations was carried out very recently by our group both at 
zero \cite{[45],[46]} 
(see also \cite{[47]}) and finite temperature \cite{[48]}. 
These 2D theoretical results 
\cite{[47],[48]} are obtained performing the regularization 
of zero-point energy 
which appears in the Gaussian fluctuations (for a comprehensive review 
in any dimension see \cite{[49]}). The zero-point energy is due to both 
fermionic single-particle excitations and bosonic collective excitations, 
and its regularization (dimensional regularization  or convergence-factor 
regularization) gives remarkable results in the full BCS-BEC crossover 
and reliable analytical prediction in the BEC regime of composite bosons, 
which are in extremely good agreement with $^6$Li experimental data 
of the zero-temperature equation of state in the crossover \cite{[41]}.

\begin{figure}[t]
\begin{center}
\includegraphics[width=.45\textwidth]{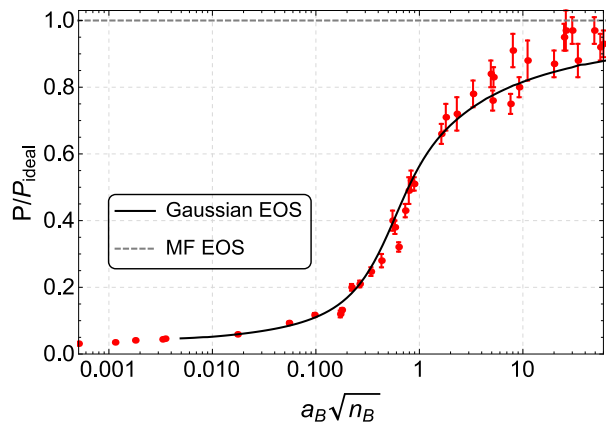}
\caption{Scaled pressure $P/(2P_{ideal})$
of the 2D gas of composite bosons as a function of the bosonic 
gas parameter $a_B n_B^{1/2}$, 
where $P_{ideal}=2\pi\hbar^2 n_B^2/m_B$ is the pressure of an
ideal 2D gas with $m_B=2m$ the mass of each bosonic particle 
(made of two fermions with mass $m$), $a_B$ is the s-wave scattering 
length of bosons, and $n_B=n/2$ is the bosonic 2D density (with $n$ 
the fermionic density). 
On the left there is the BEC regime of deeply bound Cooper pairs 
(forming bosonic molecules) while on the right there is the BCS regime 
of weakly bound Cooper pairs. The filled squares with error bars 
are the experimental data of Makhalov {\it et al.} \cite{[41]}. 
The solid curves is obtained from the zero-temperature 
beyond-mean-field theory with renormalized Gaussian fuctuations 
\cite{[46],[47],[48]}. 
Notice that the mean-field theory (dashed line) predicts 
a completely wrong result: a constant pressure, 
independent of the scattering length.}
\end{center}
\end{figure}

In 2014 Makhalov {\it et al.} \cite{[41]} have realized a quasi-2D Fermi
system with widely tunable s-wave interactions
nearly in a ground state, investigating an
ultracold gas of $^6$Li atoms by measuring the pressure $P$ as a function of
the scattering length. The experiment \cite{[41]} covers physically 
different regimes corresponding to weakly or strongly attractive
Fermi gases or a Bose gas of tightly bound pairs of fermions. 
In Fig. 1 we plot the pressure $P$ in units of the ideal pressure
$P_{ideal}$ as a function of the adimensional gas 
parameter $a_B n_B^{1/2}$, where $a_B$ is the 
scattering length of composite bosons (made of two fermions) and 
$n_B$ is the bosonic density. 
In the full crossover the figure shows a very good agreement 
between the experimental data (filled squares with error bars) and 
our theoretical results (solid curve) based on zero-temperature 
beyond-mean-field theory with renormalized Gaussian 
fluctuations \cite{[46],[47],[48]}. 

\section{Preliminar results at finite temperature} 

As previously stressed, the study of the 2D BCS-BEC crossover is very 
interesting also for high-T$_{c}$ superconductivity: the phase diagram of 
cuprate superconductors can be interpreted in terms of a BCS-BEC crossover 
as doping is varied and the critical temperature T$_c$ has a wide fluctuation 
region with pseudo-gap effects not yet fully understood \cite{[40],[50]}. 
Moreover, it has been recently suggested that iron-based superconductors 
have composite superconductivity, consisting of strong-coupling BEC in 
the electron band and weak-coupling BCS-like superconductivity in the 
hole band \cite{[51]}. 

The superfluid transition in two dimensions is of the 
Berezinskii-Kosterlitz-Thouless (BKT) type, 
featuring no true long-range coherence and making therefore 
the observation of the ordered state more subtle. 
A reliable microscopic analysis of the the BKT phase transition must 
include the quantization of circulation. This quantiziation  is a peculiar 
consequence of the existence of an underlying compact real field, whose 
spatial gradient determines the local superuid velocity of the system. 
This compact real field, the so-called Nambu-Goldstone field, is the phase 
angle of the complex bosonic field which describes, in the case of 
attractive fermions, strongly-correlated Cooper pairs of fermions with 
opposite spins \cite{[2],[45],[46],[47],[48],[49]}. 
The compactess of the Nambu-Goldstone field 
implies the presence of quantized vortices and antivortices which 
regularize the superfludid density of the system as the temperature 
increases (enhancement of thermal fluctuations) \cite{[2]}. 

By means of functional integration with renormalized 
Gaussian fluctuations we have calculated the finite-temperature 
equation of state in the 2D BCS-BEC crossover \cite{[45],[46],[47],[48],[49]}. 
We have also derived the bare superfluid density \cite{[48]} and then we have  
renormalized it by using the Kosterlitz's renormalization 
group equations \cite{[52]}: the renormalized superfluid density jumps 
discontinuously from a finite value to zero 
as the temperature reaches the BKT critical temperature. Above $T_{BKT}$ 
there is the unbinding of vortex-antivortex pairs and the 
proliferation of free vortices \cite{[2],[52]}. 

\begin{figure}[t]
\begin{center}
\includegraphics[width=.45\textwidth]{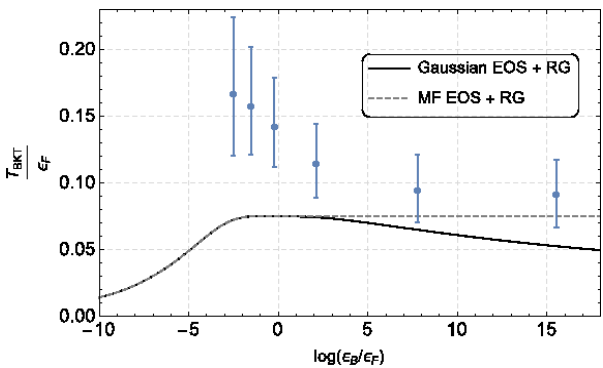}
\caption{The Berezinskii-Kosterlitz-Thouless critical temperature $T_{BKT}$ 
as a function of the bound-state binding energy $\epsilon_B$ of Cooper pairs. 
On the left there is the BCS regime characterized by a very small 
$\epsilon_B$ while on the right there is the BEC regime 
characterized by a large $\epsilon_B$. Here $\epsilon_F=\hbar^2\pi n/m$ 
is the 2D Fermi energy of the ideal two-component gas 
made of fermions with mass $m$ and number density $n$. 
The dashed line is the result of renormalization group (RG) analysis 
on top of mean-field results, whereas the solid line uses the 
Gaussian theory as the starting point for RG calculations. 
Dots with error bars are the 
experimental data of Ref. \cite{[43]}.}
\end{center}
\end{figure}

Our theoretical determination of the critical temperature $T_{BKT}$ across the 
whole crossover in reported in Fig. 2 (solid line). 
The rapid decrease of $T_{BKT}$ approaching both the BCS and 
the BEC limit is a consequence of the fermionic single-particle 
excitations and bosonic collective excitations dominating 
the superfluid density, respectively, rapidly decreasing the 
normal density as either limit is approached. A consequence of 
this interplay is that the critical temperature is higher 
in the intermediate regime ($\epsilon_B \sim \epsilon_F$), where the 
superfluid density is neither fermion-dominated nor boson-dominated.

The current approach, involving the inclusion of Gaussian fluctuations 
in the equation of state \cite{[48]} along with a renormalization group 
analysis \cite{[52]}, is able to reproduce the downward trend 
as the interaction get stronger (i.e. by increasing the binding energy 
$\epsilon_B$ of Cooper pairs). The renormalization group analysis 
on top of a mean-field theory is not sufficient to reproduce the correct trend, 
as shown by the dashed line in Fig. 2. In other words, as also 
observed elsewhere \cite{[45],[46],[47],[48],[49]}, Gaussian fluctuations 
are required in order to correctly describe the physics of an 
interacting Fermi gas in the strongly-coupled limit.

Notice that that experimental data of Fig. 2 
may be affected by errors larger than the bars. 
The algebraic decay of the first-order correlation 
function, presented in Ref. \cite{[43]} as the signature of the 
superfluid state, could be interpreted in terms of the 
strong-coupling properties of 
a normal-state, as suggested in Ref. \cite{[53]}. Moreover, 
in the experiment \cite{[43]} the fermionic 
superfluid is not trapped in a strictly 2D configuration.  

\section{Conclusions}

A complete understanding of the effects of quantum and thermal 
fluctuations on the critical temperature is of paramount importance 
for ultracold atoms physics but it is also of direct interest for 
engineering novel high-temperature superconductors. 
The results discussed here are a step towards 
a clear description of the behavior of 
the critical temperature of 2D Fermi superfluids in the BCS-BEC crossover. 
As a matter of fact, the behavior of the critical temperature of 2D Fermi 
superfluids both in the intermediate and strong-coupling regimes is still 
unknown and it is of primary importance to overcome this lack of knowledge, 
not only for ultracold atomic gases but also for high-T$_c$ 
superconductors \cite{[40],[50],[54]}. More generally, the results 
and techniques developed for superfluid alkali-metal atoms are of direct 
interest for strictly related phenomena in quite different physical 
systems, for instance high-temperature superfluidity in double-bilayer 
graphene \cite{[55]}, BCS-BEC in atomic nuclei \cite{[56]}, 
and color superconductivity in quark matter \cite{[57]}. 

\begin{acknowledgements}

This work was partially supported by MIUR through the PRIN Project "Collective 
Quantum Phenomena: from Strongly-Correlated Systems to Quantum Simulators". 

\end{acknowledgements}

\end{document}